\documentclass[preprint,nofootinbib,aps,twocolumn,superscriptaddress,showpacs]{revtex4-1}
\usepackage{color,graphicx,shortvrb,epsfig}

\usepackage{graphicx}
\usepackage{dcolumn}
\usepackage{bm}

\usepackage{colordvi}
\usepackage{epsfig,amssymb,amsmath,latexsym}

\def\lsim{\mathrel{\rlap{\lower4pt\hbox{\hskip1pt$\sim$}}
    \raise1pt\hbox{$<$}}}         
\def\gsim{\mathrel{\rlap{\lower4pt\hbox{\hskip1pt$\sim$}}
    \raise1pt\hbox{$>$}}}         

\def\lsim{\mathrel{\rlap{\lower4pt\hbox{\hskip1pt$\sim$}}
    \raise1pt\hbox{$<$}}}         
\def\gsim{\mathrel{\rlap{\lower4pt\hbox{\hskip1pt$\sim$}}
    \raise1pt\hbox{$>$}}}         

\def\beq{\begin{equation}}
\def\eeq{\end{equation}}
\def\ba{\begin{eqnarray}}
\def\ea{\end{eqnarray}}

\def\<{\langle}
\def\>{\rangle}

\begin{document}

\begin{flushright}
\date{}
\end{flushright}

\title{Common anesthetic molecules prefer to partition in liquid disorder phase domain in a composite multicomponent  membrane 
}
\
\author{Anirban Polley} 
\affiliation{Department of Chemical Engineering, Columbia University, New York City, New York-10026, USA}


\begin{abstract}
Despite a vast clinical application of anesthetics, the molecular level of understanding of general anesthesia is far from our reach. Using atomistic molecular dynamics simulation, we study the effects of common anesthetics: ethanol, chloroform and methanol in the fully hydrated symmetric multicomponent lipid bilayer membrane comprising of an unsaturated palmitoyl-oleoyl-phosphatidyl-choline (POPC), a saturated palmitoyl-sphingomyelin (PSM) and cholesterol (Chol) which exhibits phase coexistence of liquid-ordered ($l_o$) - liquid disordered ($l_d$) phase domains.
We find that the mechanical and physical properties
 such as the thickness and rigidity of the membrane are reduced while the lateral expansion of the membrane is exhibited in presence of anesthetic molecules. 
Our simulation shows both lateral and transverse heterogeneity of the anesthetics in the composite multicomponent lipid membrane.
Both ethanol and chloroform partition in the POPC-rich  $l_d$ phase domain, while methanol is distributed in both $l_o-l_d$ phase domains. 
Chloroform can penetrate deep into the membrane, 
while methanol partitions mostly at the water layer closed to the head-group and ethanol at the neck of the lipids in the membrane. 


\end{abstract}

\maketitle

\section{\label{sec:level1}Introduction}
\noindent

The influence of small molecules on the bilayer membrane has been extensively studied in last few decades. It is well known that the alcohols and chloroform are the potential  anesthetic candidates. Although anesthetics are used in every single day in all hospitals to perform painless surgical operations, the molecular level understanding of the mechanism of the general anesthesia still remains opaque.

After the discovery of the anesthetic property, chloroform is abandoned used in all hospitals until its extreme toxicity to the cells and tissues are recognized. Specifically, chloroform disrupts the normal physiological functions of the plasma membrane by its strong effect on the physical properties of the lipid membranes \cite{Meek}. On the other hand, alcohols are widely served as beverages. Here, yeast (Saccharomyces cerevisiae) is commonly used for the fermentation in the wine industry that sustains with the high ethanol concentration. However, the process of wine fermentation in the industry has been suspended dramatically around $10\%$, called `stuck-fermentation' which does not have a satisfactory understanding of the effect \cite{Block,Golovina}.

Though several theoretical and experimental studies have been engaged to investigate the path of the general anesthesia, it is still a controversial issue. In the context of the mechanism of the general anesthesia, one hypothesis is based on the influence of anesthetics on the specific proteins. It is believed that anesthetic molecules bind directly to the specific receptors in the transmembrane proteins and block the protein functions by changing its conformational equilibria \cite{Frank_1994,Frank_1997}. The other hypothesis suggests that there is an indirect mechanism involved in the anesthesia where the physical properties of lipid membrane have been changed nonspecifically by anesthetic molecules and hence, it alters the activity of the membrane proteins \cite{Regen_2009,Ueda_1998,Ueda_2001}.

The effects of anesthetics on the biological systems have been studied experimentally using the variety of different techniques, such as NMR, X-Ray, AFM imaging and theoretically which reveal that anesthetic molecules alter the lipid structures of the membrane. NMR spectroscopic studies reveal that ethanol molecules can have disordering effect on the lipids \cite{Feller_2002,Holte_1997,Barry_1995,Patra_2006,Joaquim_2011,Igor_2012}. It is also observed that ethanol molecules interact with head groups of the lipids via hydrogen bonding. The optical birefringence measurement study on the synthetic bilayer suggest that chloroform increases the acyl chain order of the lipid membrane \cite{Mishima_2003}. In differential scanning calorimeter and X-Ray studies reveal that ethanol produces structural changes in the lipid membrane above its main phase transition temperature. Partitioning of the alcohols into the water/lipid phases have been studied theoretically \cite{Cantor_1997a,Cantor_1997b,Cantor_1998}. 
 Atomistic molecular dynamics simulations have been used to elucidate the effect of anesthetic molecules on the model membrane \cite{Patra_2006,Bandyopadhyay_2004,Bandyopadhyay_2006,Smit_2004,Terama_2008,Griepernau_2007,Faller_2007,
Vierl_1994,Jackson_2007,Gawrisch_1995,Dunn_1998,Mcintosh_1984}.

In the family of small alcohol molecules, ethanol and methanol are structurally similar. Ethanol has CH3-CH2- group, while methanol has only CH3- group attached to the hydroxyl (OH-) group, respectively	. The methanol is highly poisonous to the central nervous system and might result blindness, coma and death. However, the methanol is found naturally in most of the alcoholic beverages. Any standard measure of drink (commonly known as hard liquor such as whiskey, vodka etc) contains around $40\%$  of alcohol by volume. 
The maximum tolerable concentration (MTC) of methanol in such a drink is $2\% (v/v)$ \cite{methanol_mtc} by volume though the value of the tolerance may vary depending on the health condition of an adult person. However, the natural occurrence of methanol is limited to $0.4\% (v/v)$ in $40\% (v/v)$ alcohol beverage by the European Union (EU) to provide a greater safety.  

In the present work, we study the effects of  common anesthetics such as ethanol, chloroform and toxic methanol on the symmetric model `raft membrane’ composed POPC, PSM and Chol. The motivation of choosing these 3 lipids is based on the fact that it is widely accepted that cell surface of the living cell membrane exhibits lipid-based micro-structured domains called `rafts' \cite{simons,simons_science2010,simons_nat_rev_2000} involved in a variety of cellular processes including {\em signaling} and endocytosis. The existence and nature of these functional cellular rafts have been discussed later with great details \cite{anirban_cell15,TrafficRaoMayor,raftreviews,sharma,debanjan,kripa_2012}. It is established from the studies on the properties of the raft membrane that these domains are composed of ternary lipid mixture of cholesterol and two other lipids with significantly different main transition temperature ($T_m$), exhibiting phase coexistence of $l_o$-$l_d$ domains. Thus it is important to study the effect of common anesthetics into the model raft membrane. 

The article is organized as follows: we first describe the details of the atomistic molecular dynamics simulations of the multicomponent bilayer membrane. Next, we present our main results of the spatial heterogeneity of the components, effects of anesthetics on the order parameter and lateral pressure profile. We end with some concluding remarks.

\section{\label{sec:level2}Methods}
\noindent
{\it Model membrane}\,:\, We study symmetric 3-component bilayer membrane embedded in an aqueous medium by
atomistic molecular dynamics simulations (MD) using {\it GROMACS-5.1}. 
We prepare the bilayer membrane at  $23^{\circ}$C at the relative concentration, $33.3\%$  of POPC, PSM, and  Chol, respectively.  
To the symmetric ternary bilayer membrane, we add $25\%$ of ethanol, chloroform and methanol to water layer from both sides of the symmetric ternary bilayer membrane, respectively. 
Here, all 4 multicomponent bilayer membranes have $512$ lipids in each leaflet (with a total $1024$ lipids) and $32768$ water molecules (such that the ratio of water to lipid is $32:1$) so as to completely hydrate the simulated lipid bilayer. The choice of the compositions and temperature are inspired from the ternary phase diagram at $23^{\circ}$C \cite{prieto} so that bilayer membrane exhibits $l_o$-$l_d$ phase coexistence.\\

\noindent
{\it Force fields}\,:\,We use the force field parameters for POPC, PSM, and Chol from the previous validated united-atom description \cite{anirban_jpcb12,anirban_cpl13,Tieleman-POPC,mikko}. We take the same previously used force-field parameters for the ethanol, chloroform and methanol \cite{anirban_cpl13,MikkoBPJ2006,ramon_jpcb2011,ramon_plosone2013}. The improved extended simple point charge (SPC/E) model has been used to simulate water molecules, having an extra average polarization correction to the potential energy function.    
\\

\noindent
{\it Initial configurations}\,:\,We get the initial configurations of the symmetric multicomponent bilayer membrane using {\it PACKMOL} \cite{packmol}.
For all simulation runs, we choose an initial condition where the components in each leaflet are homogeneously mixed.
\\

\noindent
{\it Choice of ensembles and equilibration}\,:\,We equilibrate the symmetric bilayers for $50$\,ps in the NVT ensemble using a Langevin thermostat to avoid bad contacts arising from steric constraints and then for $500$\,ns in the NPT ensemble ($T = 296$\,K ($23^{\circ}$C), $P =1$\,atm). The simulations are carried out in the NPT ensemble for the first $50$\,ns using Berendsen thermostat and barostat, then  using Nose-Hoover thermostat and the Parrinello-Rahman barostat to produce the correct ensemble. Each simulation has been repeated for 4 times, i.e., total $2\mu s$ simulation has been performed for a given composition of the lipid membrane. We use a semi-isotropic pressure coupling with compressibility $4.5\times 10^{-5}$ bar$^{-1}$ for the simulations in the NPT ensemble. Last $200\,ns$ of the trajectories have been used for the data analysis.

The long-range electrostatic interactions are incorporated by the reaction-field method with cut-off $r_c = 2$\,nm, while for the Lennard-Jones interactions we use a cut-off of $1$\,nm
\cite{anirban_jpcb12,mikko,patra2004}.

We calculate the lateral pressure profiles in the bilayer using Irving-Kirkwood contour and grid size $0.1$\,nm. We calculate the pairwise forces by rerunning the trajectory with cut-off  $2$\,nm for electrostatic interactions using LINCS algorithm to constrain the bond lengths \cite{Hess} and the SETTLE algorithm to keep the water molecules rigid \cite{SETTLE} so that integrator time step of $2$\,fs can be used. We generate pressure profiles from trajectories over $200$\,ns using SHAKE algorithm \cite{SHAKE} to constrain bond lengths.
\\

\noindent
{\it Computation of Voronoi Tessellation}\,:\, We perform the structural analysis of the multicomponent symmetric bilayer membrane by the use of Voronoi Tessellation. We project the position of the center of mass of the POPC, PSM, Chol and anesthetic molecules (ethanol, chloroform, and methanol) on the $x-y$ plane and Voronoi Tesselation analysis of each leaflet have been performed using algorithm available in 
MATLAB R2013b. Here, Voronoi polygon is used to calculate the area of the individual molecule in the membrane.

\begin{figure*}[h!t]
\begin{center}
\includegraphics[width=18.0cm]{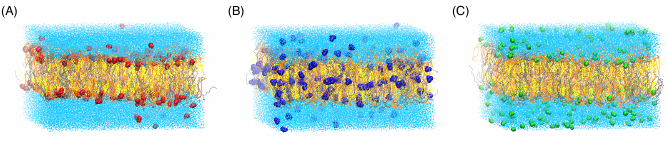}
\caption{Snapshot of equilibrium configuration of the symmetric bilayer membranes comprising POPC (gray), PSM (orange), Chol (yellow) and water (cyan) with (A) ethanol (red), (B) chloroform (blue), (C) methanol (green), respectively.  
(A) shows that ethanol partitions at the lipid-water interface, whereas chloroform can penetrate anywhere of the membrane shown in 
(B). But, methanol mostly stays in water to head group of the lipids in the membrane shown in 
(C).
}
\label{snapshot_ECM}
\end{center}
\end{figure*}

\section{\label{sec:level3}Results and Discussion}

We monitor the mean energy of the system and area per lipid throughout the simulations to ensure that the bilayer membrane reaches the chemical and thermal equilibration \cite{anirban_cpl13}.
 The last snapshot of simulations of the symmetric multicomponent bilayer membrane having ethanol, chloroform and methanol, respectively are shown in Figure\,\ref{snapshot_ECM}. It shows that ethanol molecules prefer to partition at the water-lipid interface (neck regions of the lipids) while chloroform can penetrate deep into the membrane. However, methanol molecules spread out from the water layer to head group of the lipids in the membrane.  

\subsection{Effects of anesthetics on lateral and transverse properties of the membrane}

\begin{figure*}[h!t]
\begin{center}
\includegraphics[width=18.0cm]{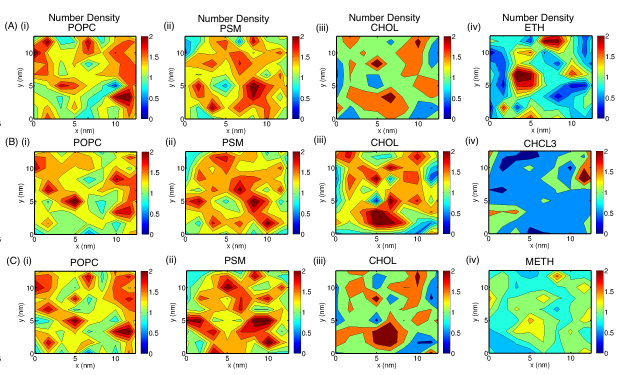}
\caption{
The spatial number density of each component in the membrane with ethanol, chloroform and methanol are shown in 
(A), (B) and (C), respectively. 
}
\label{num_density_ECM}
\end{center}
\end{figure*}

We perform simulation on symmetric bilayer membrane composed of POPC, PSM, and Chol in a ratio $1:1:1$ at $23^{\circ}$C temperature to get the phase coexistence of $l_o-l_d$ phase \cite{anirban_jpcb12}.

\begin{figure*}[h!t]
\begin{center}
\includegraphics[width=18.0cm]{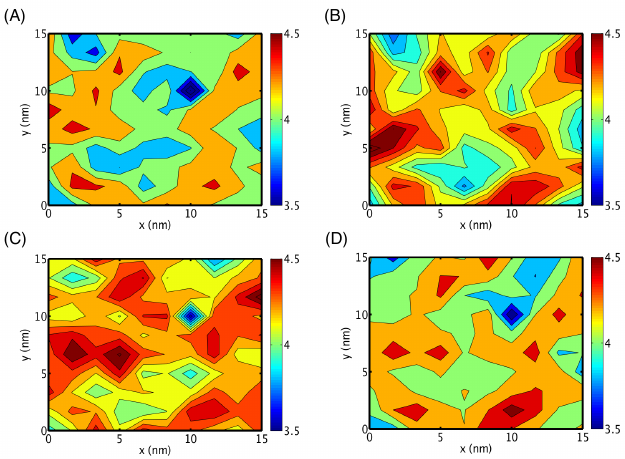}
\caption{Spatial heterogeneity of thickness of the ternary symmetric bilayer membrane comprised of POPC, PSM and Chol without and with ethanol, chloroform and methanol shown in Figure \ref{thickness} (A), (B), (C) and (D), respectively.
LUT bar indicates the thickness of the membrane in nm. 
}
\label{thickness}
\end{center}
\end{figure*}

To calculate the spatial number density, we collect the positions of the components from the last $200$\,ns trajectory of the simulations and  project the position of the center of mass of each lipid and anesthetics on the $x-y$ plane. This is then binned over a spatial scale of $1$\,nm shown in Figure\,\ref{num_density_ECM}. 
Similarly, we calculate the spatial heterogeneity of the thickness of the membrane
 by constructing the surfaces both for upper and lower leaflets separately 
  with same grid size $1$\,nm from the position of the head group of the lipids (P-atom of the POPC/PSM). The differences between the $z$-coordinates corresponding to the upper and lower surfaces with same $x-y$ coordinates gives us the spatial variation of the thickness of the bilayer membrane shown in Figure \ref{thickness}.
 


The spatial number density of each component of the bilayers with ethanol, chloroform, and methanol are shown in Figure \ref{num_density_ECM} (A), (B) and (C), respectively. The spatial number density profile shows that symmetric multicomponent bilayers having ethanol, chloroform, and methanol, respectively exhibit phase coexistence of POPC enriched $l_d$ domains and PSM with Chol enriched $l_o$ domains.
Figure \ref{num_density_ECM} (A) (iv) and (B) (iv) show that ethanol and chloroform, both are accumulated at the $l_d$ phase domains where POPC is enriched. But, the methanol is distributed homogeneously throughout the bilayer as shown in Figure \ref{num_density_ECM} (C) (iv).

The electron density ($e/nm^2$) of each component of the bilayers with ethanol, chloroform, and methanol with z-axis is shown in Figure \ref{e_density_ECM} (A), (B) and (C), respectively. Ethanol molecules are partitioned and accumulated at the neck of the lipids from both sides of the bilayer shown in Figure \ref{e_density_ECM} (A). The peaks of the electron density in Figure \ref{e_density_ECM} (B) indicates that chloroform can penetrate deep into the membrane whereas  methanols could not insert into the bilayer and therefore, spread out at the water layers  and at the head-group of the lipids of the bilayer membrane as shown in Figure \ref{e_density_ECM} (C).

As methanol stays mostly at the water layer close to the head-group of the lipids of the bilayer, it has a very minute effect on the lipid compositional heterogeneity of the bilayer. However, ethanol and chloroform can insert into the membrane and prefer to partition in the more fluidize environment of $l_d$ phase  with the floppy, loosely packed and disordered acyl chains of the POPC-rich domain.

\begin{figure*}[h!t]
\begin{center}
\includegraphics[width=18.0cm]{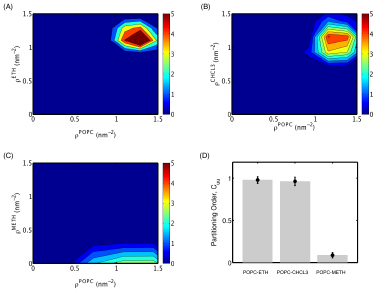}
\caption{Joint probability distribution, JPD of the number density of the POPC to that of ethanol, chloroform, and methanol, respectively are shown in (A-C). Partitioning order parameter, $C_{uu}$ defined from correlation $C(\rho^{POPC},\rho^{Anesthetics})$ between the number density of POPC to that of anesthetics are shown in (D).
}
\label{Phase_ECM}
\end{center}
\end{figure*}

\begin{figure*}[h!t]
\begin{center}
\includegraphics[width=18.0cm]{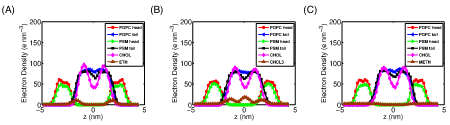}
\caption{The electron density of each component vs. z-axis (normal to membrane) of the symmetric bilayer membrane having ethanol, chloroform and methanol, respectively are shown in  (A-C).  
}
\label{e_density_ECM}
\end{center}
\end{figure*}

We have studied the extent of partitioning of the anesthetic molecules by computing the joint probability distribution, JPD of the coarse-grained number density profiles of the POPC to that of anesthetics in the same leaflet at same coarse-grained spatial location ($x,y$) from Figure \ref{num_density_ECM}. Here, we ask the question: what is the probability of partitioning of the anesthetic molecules to the POPC enriched domain in the membrane.

In Figure \ref{Phase_ECM} (A-B), the JPD shows that a distinct peak along the diagonal for the JPD of POPC-ethanol and POPC-chloroform which give the clear evidence of partitioning of ethanol and chloroform in the POPC enriched domain. But the diagonal peak in the JPD of POPC-methanol is absent as shown in \ref{Phase_ECM} (C) which signifies that methanol is not accumulated with the POPC-rich domain. 

To extent our the investigation on the spatial partitioning of the anesthetics, we define the `partitioning correlation coefficient' from the normalized cross-correlation ($r \equiv (x,y) $ and Anesthetics $\subset$ \{ethanol, chloroform \& methanol\} ), \\

\begin{widetext}
\begin{equation} \nonumber
C(\rho^{POPC}(r),\rho^{Anesthetics}(r))=\frac{\langle \rho^{POPC}(r) \rho^{Anesthetics}(r) \rangle - \langle \rho^{POPC}(r) \rangle \langle \rho^{Anesthetics}(r) \rangle}{\sqrt{\langle \rho^{POPC}(r)^2 \rangle - \langle \rho^{Anesthetics}(r) \rangle^2} \sqrt{\langle \rho^{POPC}(r)^2 \rangle - \langle \rho^{Anesthetics}(r) \rangle^2}} 
\end{equation}
\end{widetext}

averaged over space (denoted by $C_{uu}$).

We compute $C_{uu}$ for POPC-ethanol, POPC-chloroform and POPC-methanol in the respective symmetric bilayer membranes shown in Figure \ref{Phase_ECM} (D) which indicates that there is the high correlation between the POPC-rich domains and ethanol/chloroform-rich domains while the low value of the $C_{uu}$ between the POPC and methanol suggest that methanol does not correlate with the POPC-rich domains in the membrane.

\subsection{Effects of anesthetics on the order parameter}
\begin{figure*}[h!t]
\includegraphics[width=18.0cm]{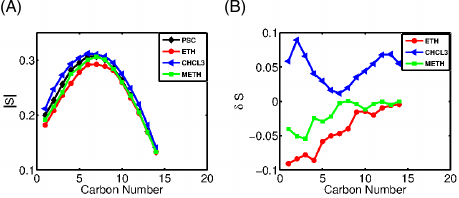}
\caption{(A) the deuterium order parameter (S) vs. carbon number of the acyl chain of the lipids of the muliticomponent symmetric membrane without (black) and with ethanol (red), chloroform (blue), and methanol(green), respectively. 
(B) the change of deuterium order parameter ($\delta s$) vs. carbon number of the acyl chain of the lipids in the membrane with anesthetics compared to without anesthetic lipid membrane.  
}
\label{scd}
\end{figure*}

To quantify the internal ordering of the lipid membranes, we compute the deuterium order parameter $S$ of the acyl chains of the lipids in the membrane and it is defined as $S=\langle \frac{3}{2}(cos^{2}\theta)-\frac{1}{2} \rangle$ where $\theta$ is the angle between the carbon atom - hydrogen (deuterium) atom and the bilayer normal. Generally, the deuterium positions are calculated from the positions of neighboring carbons assuming ideal $sp2$/$sp3$ hybridization geometries ignoring vibrational fluctuation of the deuterium atom.

\begin{figure*}[h!t]
\begin{center}
\includegraphics[width=18.0cm]{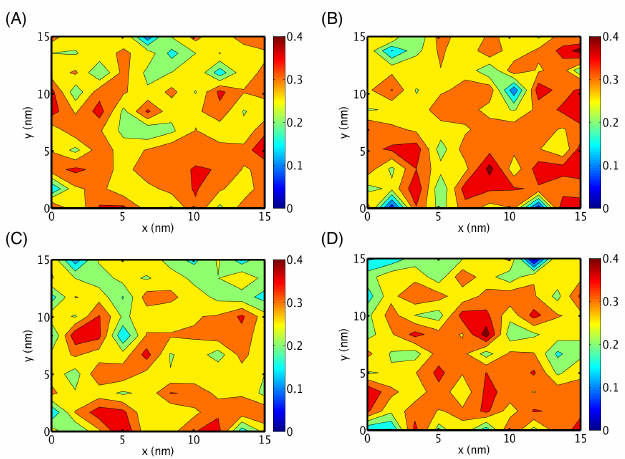}
\caption{ Spatial heterogeneity of the deuterium order parameter, $S$ of the ternary symmetric bilayer membrane comprised of POPC, PSM and Chol without and with ethanol, chloroform and methanol shown in Figure \ref{scd_spatial} (A), (B), (C) and (D), respectively. LUT bar indicates the value of S \cite{anirban_jpcb12}.
}
\label{scd_spatial}
\end{center}
\end{figure*}
We compute the average value of the $S$ corresponding to the carbon number of the acyl chains averaged over all the lipids of the symmetric bilayer Figure \ref{scd}. 
To quantify the effect of different anesthetic molecules (ethanol, chloroform and methanol) on $S$, we define, \\
$\delta S=\frac{S^{A}-S^{0}}{S^{0}}$ where $S^{A}$ and $S^{0}$ are the order parameter of the lipid chains of the symmetric multicomponent bilayer membrane {\em with} and  {\em without} anesthetic molecules, respectively.

Figure \ref{scd} shows the value of $S$ and $\delta S$ as a function of carbon numbers of the lipid chains of the membranes. We find that the value of $S$ of the lipid chains are increased in the bilayer having chloroform compared to the ternary symmetric bilayer without any anesthetic molecules whereas, that value of the lipid chains has been decreased for the bilayer with ethanol/methanol as shown in Figure \ref{scd} (A).
 In Figure \ref{scd} (B), we find that the value of $\delta S$ for the bilayer with chloroform has been increased which indicates the enhancement of the ordering of the lipids in presence of the chloroform.
Here, we find two peaks off of the $\delta S$: one at the carbon number $2$ and another carbon number $13$, respectively suggesting the abundance of the chloroform molecules at the neck and deep into the membrane as shown in  Figure \ref{e_density_ECM} (B). On the contrary, the value of $\delta S$ has been decreased for both the bilayers with ethanol/methanol. Figure \ref{scd} (B) shows that the disordering effect of the ethanol molecules is higher compared to the methanol in the bilayer membrane. Both ethanol and methanol lower the value of order parameter of the lipids chain at the neck (carbon number 3) since they are accumulated at the head/neck of the lipids of the membrane. Hence, the value of $\delta S$ is smaller at the initial number of carbon atoms of the acyl chain of the lipids in the membrane with ethanol/methanol which saturates to zeros with the higher value of carbon number of the acyl chain of the lipids.

We also compute the spatial heterogeneity of $S$ for all 4 systems to analyze the partitioning of anesthetics in the $l_o-l_d$ phase coexistence of the membrane shown in Figure \ref{scd_spatial}. To calculate the spatial heterogeneity of $S$, we calculate $S$ for selected carbon atom (C5-C7) of both POPC and PSM lipids and the center of mass of that selected atoms. Now, we divide the box with a bin size $1$\.nm and average out the value of $S$ of those lipids in the corresponding bin.

Figure \ref{scd_spatial} shows phase coexistence of ordered and disorder phase domains which is consistent with Figure \ref{thickness}.

\subsection{Crowding of lipids and anesthetics in the composite membrane}

\begin{figure*}[h!t]
\begin{center}
\includegraphics[width=18.0cm]{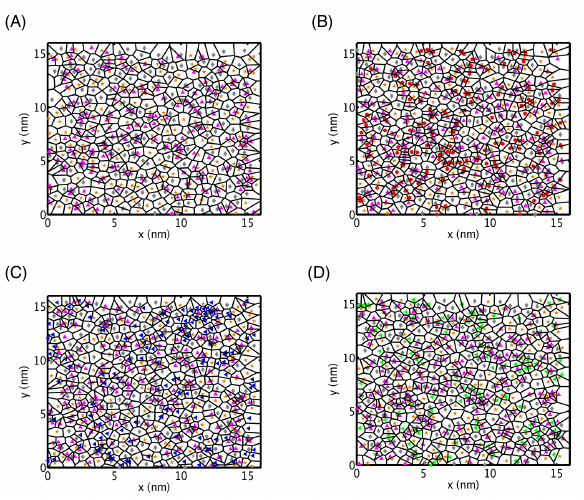}
\caption{Voronoi tessellation of a leaflet of the ternary symmetric bilayer membrane comprised of POPC (gray), PSM (orrange) and Chol (magenta) without and with ethanol (red), chloroform (blue) and methanol(green) shown in Figure \ref{voronoi} (A), (B), (C) and (D), respectively.
}
\label{voronoi}
\end{center}
\end{figure*}

\begin{figure*}[h!t]
\begin{center}
\includegraphics[width=18.0cm]{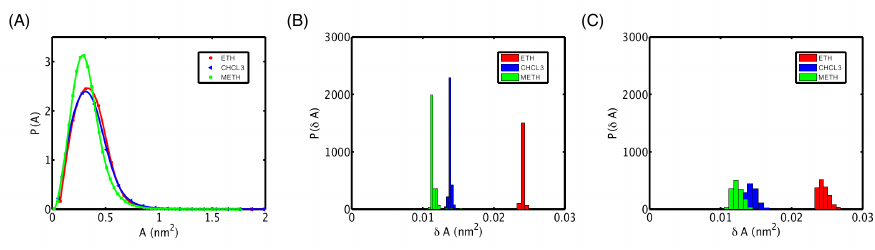}
\caption{(A) the probability distribution of the area, A of the ethanol (red), chloroform (blue), and methanol (green) in the ternary symmetric bilayer membrane computed from Figure \ref{voronoi}. 
(B) and (C) show the probability distribution of the change of the area, $P(\delta A)$ of POPC and PSM in the membrane with ethanol (red), chloroform (blue), and methanol (green) compared to without anesthetic membrane, respectively.
}
\label{area}
\end{center}
\end{figure*}

The area per lipid is one of the important properties of the lipid membranes. 
The average area per lipid of the ternary symmetric bilayer membrane and that with ethanol, chloroform and methanol are given in Table -I. It suggests that the area per lipid of the membranes have been increased in the presence of anesthetic molecules (ethanol, chloroform, and methanol).

Figure \ref{thickness} indicates that  all $4$ bilayer membranes exhibit the coexistence of thicker $l_o$ and thinner $l_d$ phase domains. Here, we tabulate the average thickness of the membrane given in Table-I which shows that the membrane becomes thinner in presence of the anesthetic molecules.

\begin{table*}[h!t]  
\begin{center}
\begin{tabular}{|c|c|c|}   \hline
System                    &   $A_h$ ($nm^2$)                &  $d$ (nm)               \\  \hline
POPC,PSM,CHOL                       &   $0.3975\pm 0.0016$            &  $4.1550\pm 0.0644$     \\  \hline
POPC,PSM,CHOL and ethanol               &   $0.4970\pm 0.0012$            &  $3.8964\pm 0.0524$     \\  \hline
POPC,PSM,CHOL and chloroform            &   $0.4512\pm 0.0013$            &  $3.7570\pm 0.0451$     \\  \hline
POPC,PSM,CHOL and methanol              &   $0.4442\pm 0.0013$            &  $3.9069\pm 0.0736$     \\  \hline
\end{tabular} 
\caption{the mean value of the area per lipid, $A_h$ and thickness, $d$ of the multicomponent bilayer membrane without and with anesthetics.}
\end{center}
\label{area_thickness}
\end{table*}

Now, we calculate the area of each lipid by Voronoi tessellation in Figure \ref{voronoi}. The Voronoi tessellation method measures the area of a molecule, which is closer to any other molecules. We use the center of mass of each molecule (lipid/anesthetics) to represent the position of the molecules and compute the area of the lipid/anesthetics of each leaflet from the ($x$,$y$) components of the center of mass of the molecules ({\em see Methods}).

The probability distribution, P($A$) of the area, $A$ of the ethanol, chloroform and methanol molecules in the bilayer membrane, respectively are shown in Figure \ref{area} (A). From the probability distribution, P($A$) (Figure 3 (A), we find that the area of the ethanol is highest whereas that of the chloroform is comparable to the methanol.

To quantify the effects of the anesthetics (ethanol/chloroform/methanol) on the area of the individual lipid (POPC/PSM) in the membrane, we define the relative increment of the area as, \\
$\delta A=\frac{A^{A}-A^{0}}{A^{0}}$ where $A^{A}$ and $A^{0}$ are the area of the lipids in the membrane {\em with } and {\em without} anesthetics, respectively. The probability distribution, P($\delta A$) of the POPC and PSM are shown in Figure \ref{area} (B) and (C), respectively. From the Figure \ref{area} (B) and (C), we find that the value of the area of the lipids (POPC/PSM) have been increased highest in the bilayer membrane having ethanol whereas, that have been increased lowest for the membrane having methanol. 
However, the increment of the area of the lipids of the membrane in presence of chloroform is very close to that of methanol. 

When we add any anesthetic molecule to the bilayer membrane, it penetrates into the bilayer. As a result, the area of the lipid has been increased for any anesthetics. Here, methanol molecules stay mostly at the water layer and head of the lipids in the membrane. Therefore, the effect of methanol in broadening the area of the lipid is least whereas, ethanol molecules stay at the neck of the lipids and push the surrounding molecules away, leading to the highest value of the area of the lipids. But, the chloroform penetrates into the membrane and can stay at any position into the membrane. As a result, the effect of chloroform in increasing the area of the lipid is slightly greater than methanol but, much less than that of ethanol. Moreover, we find that the height of the peak in the probability distribution, P($\delta A$) is much higher whereas the width of the peak in that distribution is much thinner in POPC compared to that in PSM as the anesthetic molecules like to penetrate in the POPC-rich $l_d$ phase domain compared to PSM-rich $l_o$ phase domain. Hence, the effect of anesthetics on the POPC lipids is more than that on the PSM lipids in the membrane.

\subsection{Effects of anesthetics on the lateral pressure profile}

\begin{figure*}[h!t]
\begin{center}
\includegraphics[width=10.0cm]{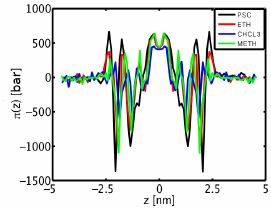}
\caption{the lateral pressure profile, $\pi(z)$ for the symmetric ternary (POPC, PSM, Chol) bilayer membrane without (black) and with anesthetic molecules (ethanol (red), chloroform (blue) and methanol (green)). 
}
\label{lateralPress}
\end{center}
\end{figure*}

The lateral pressure of the membrane is thought to be crucial in both protein activation and general anesthesia as these anesthetic molecules might affect the protein functionality by changing the lateral pressure profile of a membrane with protein embedded into it \cite{Cantor_1997a,Cantor_1997b,Cantor_1998,Gullingsrud_2004}. Though it is hard to carry out an experiment to estimate the local fluctuation of the pressure profile, few recent experiments support that small anesthetic molecule can disrupt the stability of KcsA channel by changing the lateral pressure profile of the membrane \cite{van_2004}.  
 
We calculate the local stress tensor $\sigma_{ij}$ from the virial as
$\sigma_{ij}(x,y,z)=1/v\sum_{\alpha}f_{i}^{\alpha}r_j^{\alpha}$ where $f_{i}^{\alpha}$ is the $i^{th}$ force component on the ${\alpha}^{th}$ particle due to all other particles in the system. From the local stress profile $\sigma_{ij}$, we compute the total force $F_i=\int \partial_{k}\sigma_{ik} \,dv$, its first moment (torque) $M_{ik}=\int(\partial_l \sigma_{il} x_{k} - \partial_l\sigma_{kl} x_{i}) \,dv$ to verify whether we achieve force balanced and torque balanced mechanically stable bilayer membrane \cite{anirban_cell15}.
 We further, calculate surface tension $\gamma=\int\pi(z) \,dz$ where $\pi(z)=\frac{1}{2}\left({\overline \sigma}_{xx}(z)+{\overline \sigma}_{yy}(z)\right)-{\overline \sigma}_{zz}(z)$ is the lateral pressure profile of the bilayer membrane to verify the equilibration of the system \cite{safran,Chaikin-Lubensky}.
We calculate elastic properties from the lateral pressure profile. We can write Canham-Helfrich free energy density for lipid bilayer membrane as $\Phi=\frac{1}{2}\kappa (C_1+C_2-C_0)^2+\kappa_G C_1 C_2$ where $\kappa$ and $\kappa_G$ are the bending rigidity and Gaussian bending rigidity, respectively and $C_0$ is the spontaneous curvature and $C_1$, $C_2$ are the local principal curvatures of the membrane. The bending rigidity and the Gaussian bending rigidity are connected to the lateral pressure profile as, $\kappa C_0=\int_0^{d/2} (z-\delta)\pi(z) \,dz$ and $\kappa_G=\int_0^{d/2} (z-\delta)^2\pi(z) \,dz$ where $\delta$ is the position of the neutral plane and $d$ is the thickness of the membrane.
The lateral pressure profile of the multicomponent lipid membrane comprising POPC, PSM and Chol and that with anesthetic molecules viz., ethanol, chloroform, and methanol are shown in Figure \ref{lateralPress}. The lateral pressure at the edge ($|z|> 2.5 \,nm$) of the simulation box is zero with a small fluctuation which suggests that all the membranes are well hydrated. The positive value of the lateral pressure profile indicates the pressure in the membrane whereas the negative value of that suggests the interfacial tension. In each leaflet, we find mainly three peaks in lateral pressure profile: one sharp negative peak due to the head group - tail group interface; another prominent negative peak due to the solvent - head group interface and one tall positive peak due to the positive pressure in the head group region of the membrane. At the middle, we find a pressure corresponding to the repulsive contribution of the steric interaction between the acyl chains. The magnitude of the peaks in the lateral pressure profile is decreased due to the presence of anesthetic molecules in the ternary symmetric lipid bilayer membrane. However, the pressure profile near the head groups of the lipid membrane is affected significantly by the presence of ethanol and methanol as both of them partitions at the membrane solvent interface and the lateral pressure profile of the membrane with and without ethanol/methanol are indistinguishable in the middle of the membrane as they are not present at the middle of the membrane. However, as the chloroform can easily penetrate deep into the membrane, the entire lateral pressure profile of the membrane is affected and all the peaks in the lateral pressure profile are suppressed compared to the ternary multicomponent bilayer membrane of POPC, PSM and cholesterol without any anesthetic molecules.

\begin{table*}[h!t]  
\begin{center}
\begin{tabular}{|c|c|c|}   \hline
System                    &   $\kappa C_{0}$ ($10^{-12} J/m$)      &  $\kappa_G$ ($10^{-20} J$)     \\  \hline
POPC,PSM,CHOL                        &           $-12.08363\pm 1.14181$       &  $-1.362328\pm 0.9590$         \\  \hline
POPC,PSM,CHOL and ethanol               &        $-14.7842\pm 4.22842$           & $-1.5717\pm 0.6002$            \\  \hline
POPC,PSM,CHOL and chloroform            &      $-15.77209\pm 4.02507$            & $-1.681658\pm 0.5157$          \\  \hline
POPC,PSM,CHOL and methanol              &      $-13.9220\pm 3.93532$             &    $-1.526991\pm 0.7419$       \\  \hline
\end{tabular} 
\caption{the mean value of the $\kappa C_0$ and $\kappa_G$ of the multicomponent bilayer membrane without and with anesthetics.}
\end{center}
\label{rigidity_value}
\end{table*}
To find out the effects of anesthetics on elastic properties of the bilayer membrane, we calculate the first and second moment of the lateral pressure profile which gives us two important properties: the product of bending rigidity, $\kappa$ and the spontaneous curvature,$C_0$ and the Gaussian bending rigidity, $\kappa_G$ as described above. The Table-II shows the values of $\kappa C_0$ and $\kappa_G$ of the symmetric multicomponent bilayer membrane in presence and absence of anesthetic molecules which suggests the decrease of the rigidity of the membranes in presence of anesthetic molecules. 
 
\section{Conclusion}
We have studied the effects of ethanol, chloroform, and methanol in the symmetric `raft membrane' composed of POPC, PSM and Chol using atomistic molecular dynamics. We explicitly study the main properties of the membrane that have been affected due to the presence of different anesthetics. Our main results in this study are: (i) chloroform can penetrate deep into the membrane compared to the other methanol/ethanol molecules whereas that is least for the methanol. However, all the anesthetics molecules affect the $l_d$ domains than $l_o$ domains in the raft. (ii) Ethanol and methanol both decrease the chain ordering of the lipids in the POPC-rich $l_d$ domain whereas, the chain ordering of the lipids has been increased in the presence of the chloroform. (iii) The area of the lipids has been increased in presence of all three anesthetic molecules in which the increase of the area of the lipid are exhibited most of the bilayer with ethanol and least in the bilayer with methanol. However, all the bilayers become thinner in the presence of ethanol/chloroform/methanol (iv) The peaks in the lateral pressure profile of the membrane have been relatively decreased and the pressure profile become smoother in the presence of either of ethanol, chloroform or methanol. As a result, the rigidity of all membranes has been decreased due to the presence of these anesthetic molecules.

The cell membrane consists of many different kinds of lipids and proteins and exhibits both transverse and lateral heterogeneity. In particular, we choose to study the `raft membrane' as it is believed that lateral heterogeneities of the lipid-protein exhibiting $l_o-l_d$ phase coexistence in the cell membrane take part in the several physiological activities like protein sorting, signaling processes. In the context of general anesthesia, membrane-protein interactions are induced by the addition of anesthetic molecules such as ethanol, methanol, and chloroform etc. Recent experiments also reveal that the potassium channels KcsA are dissociated due to the change of lateral pressure induced by small alcohol molecules. Here, we have characterized the simple raft membrane in the presence of anesthetic molecules. The detailed characterization presented here is essential for the simulation study of raft-protein-anesthetic systems. From our results, we find that membrane properties have been changed significantly by the anesthetic molecules, might affect the membrane-protein interaction. This is also supported by the recent experiments on enflurane-DPPC \cite{Paternostre_2003} and halothane-membrane \cite{Scharf_1998}. Though there are many open questions related to small molecules and anesthesia, our study would be useful and give a direct access to study in the context of finding the mechanism of general anesthesia.
 
\section{Acknowledgement}
AP acknowledges generous the computing facilities of `RRIHPC' clusters at RRI, Bangalore, India and Columbia University, New York. The author also thanks,
TCSC - Tampere Center for Scientific Computing resources, Finland.

\bibliographystyle{apsrev4-1}
%
%

\end{document}